\documentclass[%
superscriptaddress,
groupedaddress,
preprint,
amsmath,amssymb,
aps,
prl,
floatfix,
longbibliography
]{revtex4-2}

\usepackage{graphicx}
\usepackage{amsmath}
\usepackage{multirow}
\usepackage{color}

\usepackage{epsfig}

\begin{document}

\author{Lorenz S. Cederbaum}
\affiliation{Theoretische Chemie, Physikalisch-Chemisches Institut, Universit\"at Heidelberg, Im Neuenheimer Feld 229, Heidelberg D-69120, Germany}
\email{Lorenz.Cederbaum@pci.uni-heidelberg.de}

\author{Alexander I. Kuleff}
\affiliation{Theoretische Chemie, Physikalisch-Chemisches Institut, Universit\"at Heidelberg, Im Neuenheimer Feld 229, Heidelberg D-69120, Germany}
\email{Alexander.Kuleff@pci.uni-heidelberg.de}

\title{Stimulated emission of virtual photons: Energy transfer by light}

\date{\today}

\begin{abstract}
 
Energy-transfer processes can be viewed as being due to the emission of a virtual photon. It is demonstrated that the emission of virtual photons and thus of energy transfer is stimulated by the sheer presence of photons. We concentrate here on interatomic/intermolecular Coulombic decay (ICD) where an excited system relaxes by transferring its excess energy to a neighbor ionizing it. ICD is inactive if this excess energy is insufficiently large. However, in the presence of photons, the long-range interaction between the system and its neighbor can utilize the photon field making ICD active. The properties of this stimulated-ICD mechanism are discussed. The concept can be transferred to other scenarios. We discuss collective-ICD where two excited molecules concertedly transfer their excess energy. Also here, the presence of photons can make the process active if the sum of excess energies were insufficient to do so. Examples with typical molecules and atoms are presented to demonstrate that these stimulated processes can play a role.

\end{abstract}





\maketitle
\newpage


Intermolecular energy-transfer processes are wide spread in nature and intensely studied. If the energy of an excited molecule is transferred to bound electronic states of its neighbors, the process is called Foerster resonance energy transfer (FRET) \cite{Foerster}. There are many applications of FRET, e.g., to exciton transfer in semiconductors \cite{exciton_Fink}, and describing the first step of photosynthesis \cite{Renger_May_Kuehn,Scholes_Fleming}. Owing to energy conservation, FRET is only possible if nuclear motion is involved and implying a timescale of picoseconds or longer \cite{Renger_May_Kuehn,Scholes_Fleming}. 

We concentrate here on another, highly-efficient electronic-energy transfer
mechanism in weakly bound systems, called intermolecular Coulombic decay (ICD) \cite{Giant,ICD_Review_CR}. Here, an excited system transfers its excess energy to ionize a neighboring system. Once the excess energy is sufficiently high to ionize the neighbor, 
energy conservation is fulfilled without the need for nuclear dynamics. Thus, the excited system as well as the neighbor can be atoms or molecules and the underlying timescale is in the femtosecond regime and becomes faster the more neighbors are present \cite{ICD_Review_CR}. ICD has many applications ranging from quantum halo systems with an extreme mean separation of the atoms  \cite{Nico_He_Dimer,Exp_He_Dimer,ICD_LiHe_Anael} to quantum dots and wells \cite{ICD_QW_Nimrod,ICD_QD_Annika,ICD_QD_Nimrod}. It has been shown to be of potential importance in radiation damage and for molecules of biological interest \cite{21st_Centuary,ICD_Till_Water,ICD_Uwe_Water,ICD_Bio_Stoychev,Uwe_Bio_Review,ICD_RA_Nature,ICD_RA_Nature_Exp,Dorn_alpha_particles,Stumpf16a,Radiation_Damage_ETMD,ICD_in_unbound_pyridines,ICD_in_unbound_PANH,ICD_in_unbound_pyridines_argon}. 

We address the situation that the excess energy of the excited system is too low to ionize the neighbor. Then, of course, ICD cannot take place. We will show, however, that the sheer presence of any photon of sufficient energy $\hbar\omega$ can stimulate the ICD process. 

We consider an excited system $M^*$ which can decay radiatively to its ground state $M$. The excitation energy $\Delta=E_{_{M^*}}-E_{_M}$ is the excess energy which can be transferred in a standard-ICD process to ionize a neighbor $A$, but in our case does not suffice to do so. In many situations, $M^*$ is produced by shining light on $M$, e.g., by a laser. Consequently, additional photons are typically present together with $M^*$. To be general, we study $M^*$ and $A$ in the presence of $N_{ph}$ photons of any  energy $\hbar\omega$ such that one photon together with the excess energy suffice to ionize $A$, i.e.,  $\Delta + \hbar\omega \geq IP_A$, where $IP_A$ is the ionization potential of $A$. 

For the standard-ICD to take place, the two involved species interact via the Coulomb interaction $V_{_{MA}}$ between them. To involve a photon, we have to include the light-matter interaction $W$. The lowest possible order of perturbation theory where the presence of a photon can stimulate ICD is thus provided by second order of the overall perturbation $V_{_{TOT}} = V_{_{MA}} + W$. 

To evaluate the stimulated-ICD rate, we make use of the text-book T-matrix which obeys the following recursion relation \cite{Davydov_Book,Gottfried_Book}
\begin{align}\label{T_Mat}
	T = V_{_{TOT}} + V_{_{TOT}}(E_I - H_0 + i0^+)^{-1}T \,,  
\end{align}
where $H_0$ is the Hamiltonian of the separate participating non-interacting entities, i.e., $M$, $A$ and the photons, $E_I$ is the initial energy of the decay process, and $0^+$ is a positive infinitesimal. The rate $\Gamma_s$ of the process is   
\begin{align}\label{T_Matrix_Rate}
\Gamma_s &= 2\pi | \langle F| T |I \rangle |^2 \rho \,.
\end{align}
$\rho$ is the density of final states of the decay. $|I \rangle$ and $|F \rangle$ are the initial and final states of the process and care must be taken that the energy is conserved, i.e., $E_F = E_I$. To obtain the T-matrix to second order, it suffices to insert $T = V_{_{TOT}}$ on the right hand side of Eq.~(\ref{T_Mat}). Initially we have an excited system $M^*$, a neighbor $A$ and $N_{ph}$ photons ($|I \rangle = |M^*A,N_{ph} \rangle$) and at the end of the process $M$, an ionized neighbor together with the emitted electron $A^+$ and one photon less ($|F \rangle = |MA^+,N_{ph}-1 \rangle$). 

Of all terms of the T-matrix only two contribute to the process giving explicitly

\begin{align}\label{Formal_Rate}
 \Gamma_s =  2\pi | \langle M^*A,N_{ph}|  V_{_{MA}}\hat{R}(E_I)W 
 +  W\hat{R}(E_I)V_{_{MA}} |MA^+,N_{ph}-1 \rangle |^2 \rho\,,  \\
 \hat{R}(E_I) =  (E_I - H_0 + i0^+)^{-1}. \hspace{5cm}\nonumber
\end{align}
%
%
The initial energy is that of the excited system, neighbor and photons $E_I = E_{M^*} + E_A + N_{ph}\hbar\omega$. Energy conservation provides the total energy of the final ionized neighbor and the kinetic energy of the emitted electron $E_{A^+} = \Delta + \hbar\omega + E_A$. The kinetic energy of the emitted electron follows immediately as $\Delta + \hbar\omega - IP_A$. 

To proceed we specify the interactions $V_{_{MA}}$ and $W$. We concentrate on energy transfer at intermolecular distances $R_{_{MA}}$ between $M$ and $A$ at which no chemical bond is formed. It is illuminating to obtain an explicit expression for the stimulated-ICD rate by applying the multipole expansion of the interaction $V_{_{MA}}$ which is valid at large distances between the two. In the leading order of this expansion, including the electrons and nuclei, one obtains for neutral systems \cite{Vibr_ICD,Ensemble_Collective}:  
\begin{align}\label{Multipole_Expansion}
	V_{_{MA}} =  \frac{\vec{\hat{d}}_{_{M}}\circ\vec{\hat{d}}_{_{A}}}{R^3_{_{MA}}},
\end{align}
For charged systems, see \cite{Ensemble_Collective}.
$\vec{\hat{d}}_{_M}$ and $\vec{\hat{d}}_{_A}$ are the dipole operators of all charged particles of $M$ and $A$, respectively. For compactness we introduce the extended scalar product $\vec{\hat{d}}_{_{M}}\circ\vec{\hat{d}}_{_{A}} = \vec{\hat{d}}_{_{M}}\cdot\vec{\hat{d}}_{_{A}} -3 (\vec{u}_{_{MA}}\cdot\vec{\hat{d}}_{_{M}})(\vec{u}_{_{MA}}\cdot\vec{\hat{d}}_{_{A}})$. $\vec{u}_{_{MA}}$ is the unit vector pointing from the center of mass of $M$ to that of $A$ and $R_{_{MA}}$ is the distance between these centers. 

The interaction between charged particles and the radiation field of frequency $\omega$ and polarization $\vec{\epsilon}$ is described in the dipole approximation by the operator \cite{Davydov_Book} 
\begin{align}\label{Light_Matter_Interaction}
	W = \alpha_{\omega} \vec{\epsilon}  \cdot  [\vec{\hat{P}}_{_{M}} + \vec{\hat{P}}_{_{A}}] [\hat{a} + \hat{a}^\dagger].
\end{align}
Here $\alpha_{\omega} = \left( \frac{2 \pi \hbar}{v \omega} \right)^\frac{1}{2}$, $v$ is the volume, $\hat{a}$ is an annihilation operator of a photon of energy $\omega$, and $\vec{\hat{P}}_{_{M}} = \sum_{i} \frac{q_i}{m_i}\vec{\hat{p}}_i$, where the sum is over all particles of $M$ with charge $q_i$, mass $m_i$ and momentum $\vec{\hat{p}}_i$, and analogously for $A$. 

With the aid of the explicit expressions for the interactions $V_{_{MA}}$ and $W$ given above, one can evaluate the ICD rate by inserting completeness in front of the resolvent $\hat{R}(E_I)$ in Eq.~(\ref{Formal_Rate}). Obviously, ICD can take place after the energy of a photon is deposited in either $M^*$ or $A$. If the photon energy $\hbar\omega$ fits to resonantly excite $M^*$, the produced highly-excited system $M^{**}$ can undergo a standard-ICD with the neighbor $A$, as recently demonstrated experimentally in \cite{ICD_Two_Photon_OV}. If, on the other hand, the photon energy fits to resonantly excite the neighbor, $M^*$ and the excited neighbor $A^*$ can undergo ICD as predicted theoretically \cite{PhysRevLett.105.043004} and verified experimentally \cite{ICD_Multiply_Excited_Exp_1,ICD_Multiply_Excited_Exp_2}. In other words, absorbed photons can enable ICD in case the excess energy available in $M^*$ is insufficient to ionize the neighbor. Experimental evidence for such a laser-enabled ICD has been reported recently in \cite{ICD_Lase_Enabled_Evidence}. 

\begin{figure}
		\includegraphics[width=6cm]{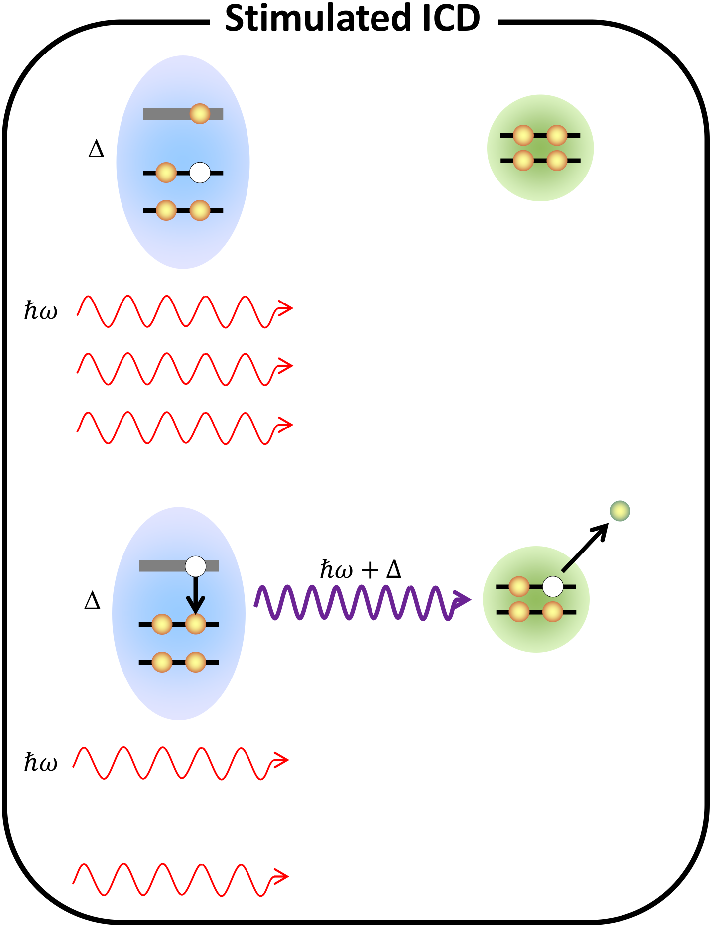}
	\caption{\label{Stimulated_emission_1} Schematic representation of the process of stimulated-ICD. If the excess energy $\Delta$ of a molecule is too low to allow for ICD to take place with a neighbor, the sheer presence of a photon of sufficient energy $\hbar \omega$ such that $\Delta + \hbar \omega$ is larger than the ionization potential of the neighbor will stimulate ICD. The rate grows with the number $N_{ph}$ of photons and $\hbar \omega$ can differ from $\Delta$. }
\end{figure}

We are not interested here in the above more obvious situations where a photon is absorbed by either the system or by its neighbor. We would like to prove that there exists a generic mechanism of stimulated-ICD applicable for all photons with sufficient energy. It is the sheer presence of the photons which matters. The stimulated emission of photons is a well-known phenomenon \cite{Davydov_Book,Gottfried_Book}. The rate in which an excited atom emits a photon in the presence of $N_{ph}$ photons of the same kind, is proportional to $N_{ph} + 1$. The sheer presence of the photons stimulates the enhanced emission. The energy transfer in ICD is usually addressed as being performed by the emission of a virtual photon \cite{Averbukh04,ICD_Review_CR,QED_Virtual_Photon} and we utilize the idea behind the stimulated emission of real photons to show that the sheer presence of photons can stimulate the emission of virtual photons. The process is schematically described in Fig.~\ref{Stimulated_emission_1}. Due to lack of sufficient excess energy, the excited system $M^*$ makes use of one of the photons present emitting a virtual photon of sufficient energy ionizing the neighbor $A$.   

To describe the ICD process due to the stimulated emission of virtual photons, we employ in the evaluation of Eq.~(\ref{Formal_Rate}) the space which can be constructed from all states of $M$ and $A$ directly involved in the process, as well as the photon states reachable. The states of $M$ and $A$ are $|M\rangle$, $|M^*\rangle$, $|A\rangle$ and $|A^+\rangle$ leading to the $8$ eigenstates of $H_0$ in Eq.~(\ref{Formal_Rate}): $|MA,n_{ph}\rangle$, $|M^*A,n_{ph}\rangle$, $|MA^+,n_{ph}\rangle$, $|M^*A^+,n_{ph}\rangle$, where $n_{ph}$ can be $N_{ph}$ and $N_{ph}-1$. This approach is named `the complete-active-space approach' since all possible states that contribute to the process under investigation are included (see also \cite{Ensemble_Collective}). Due to cancellation of terms none of the above eigenstates can be omitted in the calculations, and all are needed for a balanced treatment.

We can now calculate the stimulated-ICD rate. It is helpful to notice that \newline $i\hbar\langle M_{i},A_{j},n_{ph}|\vec{\hat{P}}_{_{M}}|M_{i'},A_{j'},n'_{ph}\rangle = (E_{M_{i'}}-E_{M_{i}})\vec{d}_{_{M_{i}M_{i'}}}\delta_{j,j'}\delta_{n,n'}$, where $\vec{d}_{_{M_{i}M_{i'}}}=\langle M_{i}|\vec{\hat{d}}_{_{M}}|M_{i'}\rangle$ is the transition dipole of the system $M$ between the indicated states. Analogous relations hold for the neighbor $A$. In the following we concentrate on the case where $A$ is an atom. Employing the complete active space to the T-matrix element in Eq.~(\ref{Formal_Rate}), we obtain:
\begin{align}\label{T_matrix_element_stimulated_ICD}
	\langle F| T |I \rangle = \frac{N_{ph}^{\frac{1}{2}} \alpha_{\omega}}{\hbar R^3_{_{MA}}}\frac{\Delta}{\Delta + \hbar\omega}[\vec{d}_{_{\delta M}}\circ \vec{d}_{_{AA^+}}][\vec{\epsilon} \cdot \vec{d}_{_{MM^*}}].
\end{align}
We remind that $\Delta=E_{_{M^*}}-E_{_M}$ is the excess energy of $M^*$. Clearly, the terms $\vec{\epsilon} \cdot \vec{d}_{_{MM^*}}$ and $\vec{d}_{_{\delta A}}\circ \vec{d}_{_{AA^+}}$ originate from the light-matter and Coulomb interactions, respectively. While $\vec{d}_{_{MM^*}}$ is the transition dipole for the ${M^*}\rightarrow{M}$ transition and $\vec{d}_{_{AA^+}}$ that for the ionizing transition ${A}\rightarrow{A^+}$, the quantity $\vec{d}_{_{\delta M}} = \vec{d}_{_{M^*M^*}} - \vec{d}_{_{MM}}$ is unusual and particularly interesting. It implies that the change of the permanent dipoles of $M$ due to the transition is a driving element in stimulated energy transfer. Another consequence of this quantity is that $M$ cannot be an atom as atoms do not have a permanent dipole.

The stimulated-ICD rate in Eq.~(\ref{T_matrix_element_stimulated_ICD}) strongly depends on the orientation of the participating systems and of the direction of the light polarization in space. For weakly bound systems, like in a cluster, the orientations of the systems are often random. It is thus illuminating to compute the rates averaged over these orientations. We choose the direction of the propagation of the light to be the $z$-axis and express all appearing vectors in Eq.~(\ref{T_matrix_element_stimulated_ICD}) accordingly. The two vectors $\vec{d}_{_{MM^*}}$ and $\vec{d}_{_{\delta M}}$ belong to the same molecule and thus have a fixed relative direction. We choose them here to be parallel to each other as is also the case in our numerical example presented below. The rate, Eq.~(\ref{T_Matrix_Rate}), is calculated using the final density-of-states of photons within a solid angle $d\Omega$ which is $d\rho = \frac{v\omega^2 d\Omega}{(2\pi c)^3 \hbar}$ \cite{Davydov_Book}, $c$ being the speed of light. Averaging over the orientation and integrating over the solid angle gives
\begin{align}\label{Stimulated_ICD_Rate_Averaged_1}
\Gamma_s = \frac{4N_{ph}}{9\hbar^3c^3R^6_{_{MA}}}	\frac{\hbar\omega \Delta^2}{(\Delta+\hbar\omega)^2} d_{_{MM^*}}^2 d_{_{\delta M}}^2 d_{_{AA^+}}^2.
\end{align}

The above formula can be made more insightful by further expressing appearing quantities by accessible experimental quantities. The photon-emission rate $\gamma_{_{M}}$ for the molecular transition $M^* \rightarrow M$ is given by the transition dipole \cite{Thorne_Book} $\gamma_{_{M}} = \frac{4\Delta^3}{3 \hbar^4 c^3} |\vec{d}_{_{MM^*}}|^2$,
and the photoionization cross section $\sigma_{_{A}}$ of the neighbor $A$ at the photon energy $\Delta + \hbar \omega$ is related to the transition dipole of $A$ into the continuum \cite{Sobelman} $\sigma_{_{A}} = \frac{4\pi^2 (\Delta + \hbar \omega)}{3c\hbar}|\vec{d}_{_{AA^+}}|^2$. The final expression for the stimulated-ICD rate now takes on the following appearance:
\begin{align}\label{Stimulated_ICD_Rate_Averaged_2}
	\Gamma_s = \frac{\hbar^2cN_{ph}}{4\pi^2R^6_{_{MA}}}	\frac{\hbar\omega}{\Delta(\Delta+\hbar\omega)^3} d_{_{\delta M}}^2 \gamma_{_{M}} \sigma_{_{A}} .
\end{align}
The excess energy $\Delta$ of ${M^*}$ is insufficient to ionize the neighbor $A$ and the presence of photons of energy $\hbar\omega$ stimulates the emission of a virtual photon of energy $\Delta+\hbar\omega$ sufficient to ionize $A$. In the presence of $N_{ph}$ photons, there are $N_{ph}$ possibilities to choose a photon for creating the required energy and the total rate is proportional to this number. Note that all quantities in the above expression for the rate can, in principle, be experimentally determined. Apart from the intermolecular distance, all quantities refer to the participating independent subsystems, the photon field, the molecule and the neighbor. 

To have a first estimate of the stimulated-ICD rate, we consider a dimer made of a pyridine molecule (C$_5$H$_5$N) and a Kr atom. Such dimers and larger clusters have been amply studied, both computationally and experimentally \cite{Molecule_rare_gas_Clusters_2,Molecule_rare_gas_Clusters_3,Molecule_rare_gas_Clusters_1}. In the dimer, Kr and pyridine are 3.6~\AA{} apart \cite{Kr_Pyridine_Dimer}. We have computed the excited states and the corresponding permanent- and transition-dipole moments of pyridine employing the extended second-order algebraic diagrammatic construction scheme [ADC(2)x] for the polarization propagator \cite{ADC2_Jochen82,ADC2x_2004}. The technical details are given in \cite{Ensemble_Collective}. We concentrate here on the $\pi\pi^*$ excitation at 7.61~eV which possesses a transition-dipole moment of 4.4~D. In the respective excited state the permanent-dipole moment is 3.3~D compared to that in the ground electronic state which is 2.2~D. Both moments point in the same direction such that $|\vec{d}_{_{\delta M}}|= 1.1$~D. The ionization of Kr requires 14.0~eV \cite{NIST} and obviously a pyridine molecule excited with 7.61~eV cannot relax via standard-ICD and ionize the neighboring krypton. The presence of a photon with energy $\hbar \omega \geq 6.39$~eV can ionize Kr via stimulated-ICD. For simplicity, we choose $\hbar \omega = \Delta = 7.61$~eV and obtain from Eq.~(\ref{Stimulated_ICD_Rate_Averaged_2}) a rate of $\Gamma_s = N_{ph}5.1 \times 10^{-11}$~eV. Converting this rate into a lifetime $\tau_s = \hbar/\Gamma_s$, one obtains $\tau_s = 12.9/N_{ph}~\mu$s. No doubt, standard-ICD is much faster than stimulated-ICD, but the latter can by no means be ignored if standard-ICD is inactive as is the case in the present example. Observing Kr${^+}$ ions after resonant excitation of pyridine in a pyridine-Kr dimer is at least a hint that it is due to stimulated-ICD. 

\begin{figure}
		\includegraphics[width=6cm]{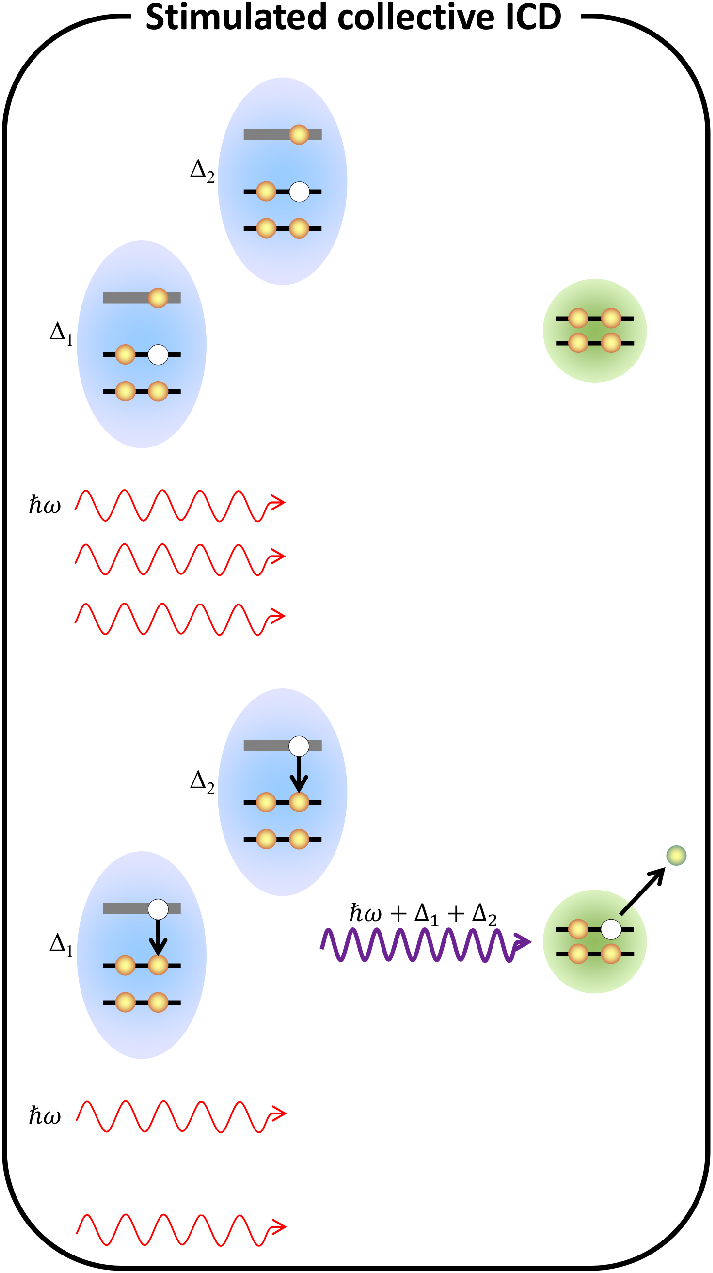}
	\caption{\label{Stimulated_emission_2} Schematic representation of the process of stimulated collective-ICD. If the sum of excess energies $\Delta_1 + \Delta_2$ of the two molecules is too low to allow for collective-ICD to take place with a neighbor, the sheer presence of a photon of sufficient energy will stimulate collective-ICD. The rate grows with the number $N_{ph}$ of photons and $\hbar \omega$ can be of arbitrary value as long as $\Delta_1 + \Delta_2 + \hbar \omega$ is larger than the ionization potential of the neighbor.}
\end{figure}

The palette of possibilities for stimulated emission of virtual photons is rather broad. We discussed above the situation where the excess energy deposited in an excited molecule is insufficient for undergoing ICD with a neighbor and participation of a single photon can lead to ICD. Depending on the photons present and the neighbor, the energy of a single or even two photons may also be insufficient to ionize this neighbor. In such cases, the above formalism can be extended to include two or three photons in order to stimulate ICD. Another, particularly promising scenario is to have two excited molecules, say $M^*_1$ and $M^*_2$. If their individual excess energies do not suffice, but their sum  suffices to ionize a neighbor $A$, collective-ICD can take place in which both molecules transfer concertedly their excess energy to $A$ \cite{Ensemble_Collective}, for atomic ions, see \cite{ICD_Collective_1}. Here, we consider the situation where the sum of excess energies of $M^*_1$ and $M^*_2$ is insufficient to allow for collective-ICD, but photons of energy $\hbar\omega$ are present such that $\Delta_1 + \Delta_2 + \hbar\omega$ is larger than $IP_A$. Will stimulated collective-ICD take place? A schematic illustration of the process is depicted in Fig.~\ref{Stimulated_emission_2}.

To answer this question, we employ tools analogous to those used above for stimulated-ICD. Now, the total perturbation $V_{_{TOT}} = V_{_{M_1A}} + V_{_{M_2A}} + V_{_{M_1M_2}} + W$  has to be used in Eq.~(\ref{T_Mat}). Initially, we have two excited molecules and an atom in its ground state and $N_{ph}$ photons are present, i.e.,  $|I \rangle = |M_1^*M_2^*A,N_{ph} \rangle$ with energy $E_I = E_{M_1^*} + E_{M_2^*} + E_A + N_{ph}\hbar\omega$. At the end of the process both molecules are in their ground state, $A$ is ionized and there is one photon less available, i.e., $|F \rangle = |M_1M_2A^+,N_{ph}-1 \rangle$ with energy $E_F = E_{M_1} + E_{M_2} + E_A^+ + (N_{ph}-1)\hbar\omega$. As the total energy is conserved, the kinetic energy of the emitted electron in this stimulated-ICD process is $\Delta_1 + \Delta_2  + \hbar\omega - IP_A$. 

Collective-ICD has been analyzed in detail in \cite{Ensemble_Collective}. If two excited molecules can relax and ionize the neighbor, second-order theory is needed. Here, since collective-ICD is energetically inactive, and the participation of a photon is required, third-order theory is needed. In analogy to Eq.~(\ref{Formal_Rate}) one finds
\begin{widetext}
\begin{align}\label{Formal_Rate_Collective}
	\Gamma_{sc} = 2\pi | \langle I| V\hat{R}(E_I)V\hat{R}(E_I)W + V\hat{R}(E_I)W\hat{R}(E_I)V + W\hat{R}(E_I)V\hat{R}(E_I)V|F \rangle |^2 \rho\,,
\end{align}
\end{widetext}
where $V = V_{_{M_1A}} + V_{_{M_2A}} + V_{_{M_1M_2}}$ includes all Coulomb-interaction terms.

In complete analogy to stimulated-ICD, we compute the rate $\Gamma_{sc}$ for stimulated collective-ICD for distances $R_{_{M_1A}}$, $R_{_{M_2A}}$ and $R_{_{M_1M_2}}$ between all pairs of the participating species at which no chemical bond is formed by employing the multipole expansion in Eq.~(\ref{Multipole_Expansion}) and the complete-active-space approach. This configuration space includes all the configurations which can be constructed from all states of $M_1$, $M_2$ and $A$ directly involved in the process, as well as the photon states reachable. The configurations to be inserted into Eq.~(\ref{Formal_Rate_Collective}) are: $|M_1M_2A,n_{ph}\rangle$, $|M_1^*M_2,A,n_{ph}\rangle$, $|M_1M_2^*A,n_{ph}\rangle$, $|M_1^*M_2^*A,n_{ph}\rangle$, $|M_1M_2A^+,n_{ph}\rangle$, $|M_1^*M_2,A^+,n_{ph}\rangle$, $|M_1M_2^*A^+,n_{ph}\rangle$ and $|M_1^*M_2^*A^+,n_{ph}\rangle$, where $n_{ph}$ can be $N_{ph}$ and $N_{ph}-1$. Since there are three terms in Eq.~(\ref{Formal_Rate_Collective}) and these configurations have to be inserted twice in front of the resolvents $\hat{R}(E_I)$, one formally encounters $3\times16^2$ terms. In addition, the Coulomb interaction $V$ now contains three terms, making the evaluation of the rate rather cumbersome, but still straightforward. We have, nevertheless, derived all terms and obtained a lengthy expression for $\Gamma_{sc}$, the general discussion of which is beyond the scope of this work. We concentrate here on the interesting situation where the two excited molecules are at a distance from each other like in a cluster and the atom $A$ to be ionized is far from them. In this case, all terms but one scale as $1/R_{_{M_1A}}^6$ or $1/R_{_{M_2A}}^6$ and can be neglected. There is a single term contributing to $\Gamma_{sc}$ which does not depend on the distance of the molecules to the atom. For transparency we consider the two molecules to be identical, i.e., $\Delta_1=\Delta_2=\Delta$, $|\vec{d}_{_{M_1M_1^*}}|^2 = |\vec{d}_{_{M_2M_2^*}}|^2 = |\vec{d}_{_{MM^*}}|^2$ and likewise also for the permanent dipole moments. Making use of the conversion of the transition-dipole moment $|\vec{d}_{_{MM^*}}|^2$ to the measurable photon-emission rate $\gamma_{_{M}}$, we obtain after averaging over the random orientation of the participating molecules
\begin{align}\label{Stimulated_Collective_ICD_Rate_Averaged}
	\Gamma_{sc} = \frac{\eta9\hbar^6c^4N_{ph}}{4^3\pi^3R^{12}_{_{M_1M_2}}}	\frac{\hbar\omega(2\Delta+\hbar\omega)}{\Delta^{10}} (d_{_{MM}}^2 + d_{_{M^*M^*}}^2)^2\gamma_{_{M}}^2 \sigma_{_{A}} 
\end{align}
for the rate of stimulated collective-ICD, where $\eta=1/16$. As the energy of the virtual photon available to ionize $A$ is $2\Delta+\hbar \omega$, the respective photoionization cross section $\sigma_{_{A}}$ is at this energy, i.e., $\sigma_{_{A}} = \frac{4\pi^2 (2\Delta + \hbar \omega)}{3c\hbar}|\vec{d}_{_{AA^+}}|^2$ \cite{Sobelman}. It is relevant to note that one can achieve an `ideal' case by choosing the transition dipoles of both molecules to be parallel to the unit vector $\vec{u}_{_{M_1M_2}}$ connecting the centers of mass of these molecules. In this case, the same expression for the rate holds, but with $\eta=1$. 

To learn about the efficiency of stimulated collective-ICD, we consider pyridine and Ar. Two excited pyridine molecules in a pyridine cluster can undergo collective-ICD with Kr at the excitation energy of 7.61~eV. We, therefore, choose Ar whose IP is 15.76~eV \cite{NIST}. The presence of a photon of 0.54~eV or more can lead to the ionization of Ar via stimulated collective-ICD. Choosing a photon of the same energy as the excitation energy and a distance between the pyridine molecules of 3.32~\AA{} \cite{pyridine_distance} gives via Eq.~(\ref{Stimulated_Collective_ICD_Rate_Averaged}) the rate per photon of  $\Gamma_{sc}/N_{ph}=8.1 \times 10^{-12}$~eV. This rate implies a lifetime $\tau_{sc} = \hbar/\Gamma_{sc} = 81.1/N_{ph}~\mu$s. This lifetime is longer than that of stimulated-ICD discussed above. In contrast to the latter mechanism where the neighbor has to be rather close to the excited molecule, this is not the case in the stimulated concerted-ICD case. As long as the participating two excited molecules are at a reasonable distance from each other, one can exploit this fact and achieve substantial rates $\Gamma_{sc}$ by having many neighbors around. The total rate is proportional to the number of these neighbors within a distance where the dipole approximation, which is the only additional approximation used (see text in front of Eq.~(\ref{Light_Matter_Interaction})), holds.  Let's consider an example. Small molecular clusters can be embedded in atomic clusters. To estimate the number $n_{Ar}$ of Ar atoms within a cluster of radius $r$, one can use the Hagena relation $n_{Ar}=\frac{2\pi}{3}\left( \frac{2r}{a_0}\right)^3$ where $a_0$ is the lattice parameter of the homogeneous cluster \cite{Cluster_Size_1}. The wavelength of the 7.61~eV photon is $\lambda=163$~nm and choosing $r=\lambda$ and $a_0=5.46$~\AA{} \cite{Crystallographic_Data} provides $n_{Ar}=0.45 \times 10^9$, giving rise to a total rate of stimulated collective-ICD per photon which is in the fs regime, i.e., in the same time range as standard-ICD in mixed clusters. 

In the common stimulated emission of photons the sheer presence of photons of the same kind enhances the emission rate. This is also the case in energy-transfer processes like ICD which are viewed as being triggered by the emission of virtual photons. There are, however, basic differences. In stimulated-ICD the photons do not have to match the excess energy of the relaxing system and, in principle, they can even be all of different energies as long as the sum of excess energy and a photon suffice to ionize the neighbor. In the absence of photons, the emission rate of an excited isolated system is due to spontaneous emission. In case of standard-ICD, the process can only take place if the excess energy is large enough to ionize the neighbor. Then, the presence of photons will typically lead to a small enhancement of ICD. The major impact of stimulated-ICD is when standard-ICD is not operative due to insufficient excess energy and ICD becomes operative only due to the sheer presence of photons. Then, ions and electrons appear as products. 

There are many scenarios of energy transfer by stimulated emission of virtual photons including energy transfer to bound states of the neighbor. We have concentrated here on ICD and also on collective-ICD where two (or more) excited molecules concertedly give rise to the ionization. These two mechanisms exhibit different characteristics and expected to become relevant in different contexts. We presented explicit examples where excited pyridine molecules and rare-gas neighbors are involved. Since the input data for these systems are typical for many systems, the results of these examples are to be seen as indications that the stimulated processes cannot be neglected and we hope that the work will stimulate further theoretical and experimental studies.

Finally, we would like to mention two points. QED formalism provides promising approaches to investigate energy transfer both via FRET \cite{Review_QED_Bradshaw,Three_body_QED_Andrews} and ICD \cite{Retardation_ICD_1}.  Such approaches also incorporate the effect of retardation not included in the present first investigation. Retardation makes the impact of energy transfer more long range. However, for the transfer energies considered in the present work the impact of retardation is very small and its incorporation is left to future studies. We stress that for standard-ICD at distances as in clusters, the rate can be much larger in reality than predicted by the asymptotic formula \cite{Averbukh04} which can be viewed as a lower limit. The true power of ICD is indeed to be seen at such distances. We expect a similar behavior also for stimulated-ICD.

\begin{acknowledgments}
	The authors thank J. Fedyk, J. Kiefka, J. Schirmer, N. Sisourat, and A. Vibok for valuable contributions. Financial support by the German Science Foundation (DFG; Grant No. CE~10/56-1) is gratefully acknowledged.
\end{acknowledgments}

\bibliography{biblio}

\end{document}